\documentclass[preprint,12pt]{aastex}
\renewcommand\bv{{\bf v}}
\renewcommand\>{{\rangle}}
\newcommand\bnabla{{\bf \nabla}}
\newcommand\bk{{\bf k}}

\newcommand\ex{\hat{\bf e}_x}
\newcommand\ey{\hat{\bf e}_y}
\newcommand\ez{\hat{\bf e}_z}
\newcommand\tco{{\tau_{co}}}
\newcommand\tc{{\tau_c}}
\newcommand\tce{{\<\< \tau_c \>\>}}
\newcommand\bO{{\bf \Omega}}
\newcommand\del{\partial}
\newcommand\<{{\langle}}

\newcommand\qq{{3\over{2}}}

\newcommand\cm{{\rm\,cm}}

\newcommand\pc{{\rm\,pc}}

\newcommand\yr{{\rm\,yr}}
\newcommand\gm{{\rm\,g}}
\newcommand\kms{{\rm\,km\,s^{-1}}}
\newcommand\gcm{{\rm\,g\,cm^{-3}}}

\newcommand\msun{{\rm\,M_\odot}}

\newcommand\K{{\rm\,K}}
\shortauthors{Johnson \& Gammie}
\shorttitle{Gravitational Instability}
\begin{document}

\title{Nonlinear Outcome of Gravitational Instability in
Disks with Realistic Cooling}

\author{Bryan M. Johnson and Charles F. Gammie}

\affil{Center for Theoretical Astrophysics,
University of Illinois at Urbana-Champaign,
1110 West Green St., Urbana, IL 61801}

\begin{abstract}

We consider the nonlinear outcome of gravitational instability in
optically-thick disks with a realistic cooling function.  We use a
numerical model that is local, razor-thin, and unmagnetized.  External
illumination is ignored.  Cooling is calculated from a one-zone model
using analytic fits to low temperature Rosseland mean opacities.  The
model has two parameters: the initial surface density $\Sigma_o$ and
the rotation frequency $\Omega$.  We survey the parameter space and find: (1)
The disk fragments when $\tce \Omega \sim 1$, where $\tce$ is an
effective cooling time defined as the average internal energy of the model
divided by the average cooling rate.  This is consistent with earlier
results that used a simplified cooling function.  (2) The initial
cooling time $\tco$ for a uniform disk with $Q = 1$ can differ by orders
of magnitude from $\tce$ in the nonlinear outcome.  The difference is
caused by sharp variations in the opacity with temperature.  The
condition $\tco \Omega \sim 1$ therefore does not necessarily indicate
where fragmentation will occur.  (3) The largest difference
between $\tce$ and $\tco$ is near the opacity gap, where dust is absent
and hydrogen is largely molecular.  (4) In the limit of strong
illumination the disk is isothermal; we find that an isothermal version
of our model fragments for $Q \lesssim 1.4$.  Finally, we discuss some physical
processes not included in our model, and find that most are likely to
make disks more susceptible to fragmentation.  We conclude that disks
with $\tce\Omega \lesssim 1$ do not exist.

\end{abstract}


\section{Introduction}

The outer regions of accretion disks in both active galactic nuclei
(AGN) and young stellar objects (YSO) are close to gravitational
instability (for a review see, for AGN: \citealt{shlos90}; YSOs:
\citealt{al93}).  Gravitational instability can be of central importance in
disk evolution.  In some disks, it leads to the efficient redistribution
of mass and angular momentum (e.g.  \citealt{lar84,lr96,gam01}).  In
other disks, gravitational instability leads to fragmentation and the
formation of bound objects.  This may cause the truncation of
circumnuclear disks \citep{goo03}, or the formation of planets (e.g.
\citealt{bos97}, and references therein).

We will restrict attention to disks whose potential is dominated by the
central object, and whose rotation curve is therefore approximately
Keplerian.  Gravitational instability to axisymmetric perturbations sets in when the sound
speed $c_s$, the rotation frequency $\Omega$, and the surface density
$\Sigma$ satisfy
\begin{equation}\label{QDEF}
Q \equiv {c_s\Omega\over{\pi G\Sigma}} < Q_{crit} \simeq 1
\end{equation}
\citep{toom64,glb}.  Here $Q_{crit} = 1$ for a ``razor-thin'' (two-
dimensional) fluid disk model of the sort we will consider below, and
$Q_{crit} = 0.676$ for a finite-thickness isothermal disk \citep{glb}. \footnote{For global 
models with radial structure, nonaxisymmetric instabilities typically set in for
slightly larger values of $Q$ (see \citealt{bos98} and references therein).}
The instability condition (\ref{QDEF}) can be rewritten, for a disk with
scale height $H \simeq c_s/\Omega$, around a central object of mass
$M_*$,
\begin{equation}\label{THICKCRIT}
M_{disk} \gtrsim {H\over{r}} M_*,
\end{equation}
where $M_{disk} = \pi r^2 \Sigma$.  For YSO disks $H/r \sim 0.1$ and
thus a massive disk is required for instability.  AGN disks are expected
to be much thinner.  The instability condition can be rewritten in a
third, useful form if we assume that the disk is in a steady state and its
evolution is controlled by internal (``viscous'') transport of angular
momentum.  Then the accretion rate $\dot{M} = 3\pi \alpha c_s^2
\Sigma/\Omega$, where $\alpha \lesssim 1$ is the usual dimensionless
viscosity of \cite{ss73}, and
\begin{equation}\label{INSTCRIT}
\dot{M} \gtrsim {3 \alpha c_s^3\over{G}}
= 7.1 \times 10^{-4}\, \alpha
\left( {c_s \over{ 1\kms}}\right)^3 \,\msun \yr^{-1}
\end{equation}
implies gravitational instability (e.g. \cite{shlos90}).  Disks
dominated by external torques (e.g. a magnetohydrodynamic [MHD] wind)
can have higher accretion rates (but not arbitrarily higher; see
\citealt{goo03}) while avoiding gravitational instability.

For a young, solar-mass star accreting from a disk with $\alpha =
10^{-2} $ at $10^{-6} \msun \yr^{-1}$, equation (\ref{INSTCRIT}) implies
that instability occurs where the temperature drops below $17 \K$.
Disks may not be this cold if the star is located in a warm molecular
cloud where the ambient temperature is greater than $17 \K$, or if the
disk is bathed in scattered infrared light from the central star
(although there is some evidence for such low temperatures in the solar
nebula, e.g. \citealt{owen99}).  If the vertically-averaged value of
$\alpha$ is small and internal dissipation is confined to surface
layers, as in the layered accretion model of \cite{gam96}, then
instability can occur at higher temperatures, although equation
(\ref{THICKCRIT}) still requires that the disk be massive.

AGN disk heating is typically dominated by illumination from a central
source.  The temperature then depends on the shape of the disk.  If the
disk is flat or shadowed, however, and transport is dominated by
internal torques, one can apply equation (\ref{INSTCRIT}).  For example,
in the nucleus of NGC 4258 \citep{miy95} the accretion rate may be as
large as $10^{-2} \msun \yr^{-1}$ \citep{las96,gam99}.  Equation
(\ref{INSTCRIT}) then implies that instability sets in where $T < 10^4
(\alpha/10^{-2}) \K$.  If the disk is illumination-dominated then $Q$
fluctuates with the luminosity of the central source.

In a previous paper \citep{gam01}, one of us investigated the effect of
gravitational instability in cooling, gaseous disks in a local model.  A
simplified cooling function $\Lambda$ was employed in these simulations,
with a fixed cooling time $\tco$:
\begin{equation}
\Lambda = -{U\over{\tco}},
\end{equation}
where $U \equiv$ the internal energy per unit area.  Disk fragmentation
was observed for $\tco\Omega \lesssim 3$.  The purpose of this
paper is to investigate gravitational instability in a local model with
more realistic cooling.  

Several recent numerical experiments have included cooling, as opposed
to isothermal or adiabatic evolution, and we can ask whether these
results are consistent with \cite{gam01}.  \cite{nel00} studied a global
two-dimensional (thin) SPH model in which the vertical density and
temperature structure is calculated self-consistently and each particle
radiates as a blackbody at the surface of the disk.  The initial
conditions at a radius corresponding to the minimum initial value of Q
($\sim 1.5$) for these simulations were $\Sigma_o \approx 50
{\rm\,g\,cm^{-2}}, \Omega \approx 8 \times 10^{-10} {\rm \,s^{-1}}$; the
initial cooling time under these circumstances is $\tco \approx 250 \,
\Omega^{-1}$, so fragmentation is not expected and is not observed.

\cite{dur01} consider a global three dimensional (3D) Eulerian
hydrodynamics model in which the volumetric cooling rate varies with
height above the midplane so as to preserve an isentropic vertical
structure.  The cooling time is fixed at each radius.  Their cooling
time $\gtrsim 10 \Omega^{-1}$ at all radii, so fragmentation is not expected
based on the criterion of \cite{gam01}.  The simulations show structure
formation due to gravitational instabilities but not fragmentation.

\cite{ric03} consider a global 3D SPH model with a cooling time that is
a fixed multiple of $\Omega^{-1}(r)$.  They find that their disk
fragments when $\tco \approx 3 \Omega^{-1}$ and $M_{disk} = 0.1 M_*$.
For a more massive disk ($M_{disk} = 0.25 M_*$), fragmentation occurred
at somewhat higher cooling times ($\tco \approx 10 \Omega^{-1}$).  This
is effectively a global generalization of the local model problem
considered by \cite{gam01}.  The fact that the results are so consistent
suggests that the local, thin approximation used in \cite{gam01} and
here give a reasonable approximation to a global outcome.

\cite{may02} consider a global three dimensional SPH model of a
circumstellar disk.  Explicit cooling is not included, but the equation
of state switches from isothermal to adiabatic when gravitational
instability begins to set in.  This is designed to account for the
inefficient cooling of dense, optically thick regions.  Fragmentation is
observed.  Realistic cooling can have a complex influence on disk
evolution, and it is not clear that switching between isothermal and
adiabatic behavior ``brackets'' the outcomes that might be obtained when
full cooling is used.

Other notable recent work, such as that by \cite{bos02}, includes strong
radiative heating in the sense that the effective temperature of the
external radiation field $T_{irr}$ is comparable to or larger than the
disk midplane temperature $T_c$.  In the limit that $T_{irr} \ll T_c$ we
recover the limit considered here and in \cite{gam01}; in the limit that
$T_{irr} \gg T_c$ the disk is effectively isothermal.

The plan of this paper is as follows.  In \S 2 we describe the model,
with a detailed description of the cooling function given in \S 3. The
results of numerical experiments are described in \S 4.  Conclusions are
given in \S 5.

\section{Model}

The model we use here is identical to that used in \cite{gam01} in every
respect except that we use a more complicated cooling function.  To make
the description more self-contained, we summarize the basic equations of
the model here.  The model is local, in the sense that it considers a
region of size $L$ where $L/r_o \ll 1$ and $r_o$ is a fiducial radius.  We
use a {\it local Cartesian} coordinate system $x \equiv r - r_o$ and $y
\equiv (\phi - \Omega t) r_o$, where $r,\phi$ are the usual cylindrical
coordinates and $\Omega$ is the orbital frequency at $r_o$.  The model is also
thin in the sense that matter is confined entirely to the plane of the disk.

Using the local approximation one can perform a formal expansion of the
equations of motion in the small parameter $L/r_o$.  The resulting equations of
motion read, where $\bv$ is the velocity, $P$ is the (two-dimensional)
pressure, and $\phi$ is the gravitational potential with the time-steady
axisymmetric component removed:
\begin{equation}
{D\bv\over{D t}} = -{\bnabla P\over{\Sigma}} - 2\bO\ez\times\bv
        + 3\Omega^2 x \ex - \bnabla\phi.
\end{equation}
For constant pressure and surface density, $\bv = -{3\over{2}}\Omega x
\ey$ is an equilibrium solution to the equations of motion.  This linear
shear flow is the manifestation of differential rotation in the local
model.

The equation of state is
\begin{equation}
P = (\gamma - 1) U,
\end{equation}
where $P$ is the two-dimensional pressure and $U$ the two-dimensional
internal energy.  The two-dimensional (2D) adiabatic index $\gamma$ can
be mapped to a 3D adiabatic index $\Gamma$ in the low-frequency (static)
limit.  For a non-self-gravitating disk $\gamma = (3\Gamma - 1)/(\Gamma +
1)$ (e.g.  \citealt{ggn86,ost92}).  For
a strongly self-gravitating disk, one can show that $\gamma = 3 - 2/\Gamma$.  We
adopt $\Gamma = 7/5$ throughout, which yields $\gamma = 11/7$.

The internal energy equation is
\begin{equation}
{\del U\over{\del t}} + \nabla \cdot (U \bv) =
	-P\bnabla\cdot\bv - \Lambda,
\end{equation}
where $\Lambda = \Lambda(\Sigma,U,\Omega)$ is the cooling function,
fully described below.  Notice that there is no heating term; heating is
due solely to shocks.  Numerically, entropy is increased by artificial
viscosity in shocks.

The gravitational potential is determined by the razor-thin disk Poisson
equation:
\begin{equation}
\nabla^2\phi = 4\pi G \Sigma \, \delta(z).
\end{equation}
For a single Fourier component of the surface density $\Sigma_{\bk}$
this has the solution
\begin{equation}
\phi = -{2\pi G\over{|\bk|}} \Sigma_{\bk} e^{i \bk\cdot{\bf x}
	- |k z|}.
\end{equation}
A finite-thickness disk has weaker self-gravity, but this does not
qualitatively change the dynamics of the disk in linear theory
\citep{glb}.

We integrate the governing equations using a self-gravitating
hydrodynamics code based on ZEUS \citep{sn92}. ZEUS is a
time-explicit, operator-split, finite-difference method on a staggered
mesh.  It uses an artificial viscosity to capture shocks.  Our
implementation has been tested on standard linear and nonlinear
problems, such as sound waves and shock tubes.  We use the ``shearing
box'' boundary conditions, described in detail by \cite{hgb1}, and solve
the Poisson equation using the Fourier transform method, modified for
the shearing box boundary conditions.  See \cite{gam01}  for further
details on boundary conditions, numerical methods and tests.

The numerical model is always integrated in a region of size $L \times
L$ at a numerical resolution of $N \times N$.  In linear theory the disk
is most responsive at the critical wavelength $2c_s^2/G\Sigma_o$.\footnote{The
wavelength corresponding to the minimum in the dispersion
relation for axisymmetric waves.}
We have checked the dependence of the outcome on $L$ and
found that as long as $L \gtrsim 2c_s^2/G\Sigma_o$ the outcome does
not depend on $L$.  We have also checked the dependence of the outcome
on $N$ and found that the outcome is
insensitive to $N$, at least for the models with $N \geq 256$ that we use.

\section{Cooling Function}\label{PCFUNC}

Our cooling function is determined from a one-zone model for the
vertical structure of the disk.  The disk cools at a rate per unit area
\begin{equation}
\Lambda \equiv 2 \sigma T_e^4,
\end{equation}
which defines the effective temperature $T_e$.  The cooling function
depends on the heat content of the disk and how that content is
transported from the disk interior to the surface: by radiation,
convection, or perhaps some more exotic form of turbulent transport such
as MHD waves.  Low temperature disks are expected to be convectively
unstable (e.g.  \citealt{cam78,lp80}).  \cite{cas93} has argued,
however, that the radiative heat flux in an adiabatically-stratified
disk is comparable to the heat dissipated by turbulence (in an
$\alpha$-disk model), suggesting that convection is incapable of
radically altering the vertical structure of the disk.  We will consider
only radiative transport.

If the disk is optically thick in the Rosseland mean sense, so that
radiative transport can be treated in the diffusion approximation, then
\citep{hub90}
\begin{equation}\label{TEFF}
T_e^4 = \frac{8}{3}\frac{T_c^4}{\tau} 
\end{equation}
where $\tau$ is the Rosseland mean optical depth and $T_c$ is the
central temperature.  We will assume that $T_c \approx T$, where
\begin{equation}\label{TEMP}
T = \frac{\mu m_p c_s^2}{\gamma k_B},
\end{equation}
and
\begin{equation}\label{CS}
c_s^2 = \gamma(\gamma - 1)\frac{U}{\Sigma},
\end{equation}
which follows from the equation of state and the assumption that the
radiation pressure is small (we have verified that this is never
seriously violated).  Here $k_B$ is Boltzmann's constant, $m_p$ is the
proton mass, and $\mu$ is the mean mass per particle, which we have set
to $2.4$ in models with initial temperature below the boundary between
the grain-evaporation opacity and molecular opacity and $\mu = 0.6$ in
models with initial temperature above the boundary.

The optical depth is
\begin{equation}
\tau \equiv \int_0^\infty \, dz \, \kappa(\rho_z, T_z) \rho_z
\end{equation}
where $\kappa$ is the Rosseland mean opacity, $\rho_z$ and $T_z$ are
local density and temperature, and $z$ is the height above the midplane.
Following the usual one-zone approximation,
\begin{equation}
\int_0^\infty \, dz \, \kappa(\rho_z, T_z) \rho_z \approx
H \kappa(\bar{\rho},\bar{T}) \bar{\rho}
\end{equation}
where the overbar indicates a suitable average and $H \approx
c_s(T)/\Omega$ is the disk scale height (we ignore the effects of
self-gravity on the disk scale height, which is valid when locally $Q
\gtrsim 1$).  Taking $\bar{T} \approx T$ and $\bar{\rho} \approx
\Sigma/(2 H)$ then gives a final, closed expression for $\Lambda$.

We have adopted the analytic approximations to the opacities provided by
\cite{bel94}.  These opacities are dominated by, in order of increasing
temperature: grains with ice mantles, grains without ice mantles,
molecules, H$^-$ scattering, bound-free/free-free absorption and
electron scattering. The molecular opacity regime is commonly called the
{\it opacity gap}; it is too hot for dust, but too cold for H$^-$
scattering to contribute much opacity.  The opacity can be as much as
$4$ orders of magnitude smaller than the $\sim 5 \gm \cm^{-2}$ typical
of the dust-dominated opacity regime.  It turns out that this feature
plays a significant role in the evolution of gravitationally-unstable
disks.

To sum up, the cooling function is
\begin{equation}\label{COOL}
\Lambda(\Sigma,U,\Omega) = \frac{16}{3} \frac{\sigma T^4}{\tau}.
\end{equation}
For a power-law opacity of the form $\kappa = \kappa_0 \rho^a T^b$, this
implies that
\begin{equation}
\Lambda \sim \Sigma^{-5 - 3 a/2 + b} U^{4 + a/2 - b}.
\end{equation}
From this it follows that the cooling time $\tc \equiv U/\Lambda$
scales as
\begin{equation}
\tc \sim \Sigma^{5 + 3 a/2 - b} U^{-3 - a/2 + b}.
\end{equation}
If the disk evolves quasi-adiabatically (as it does if the cooling time
is long compared to the dynamical time) then $U \sim \Sigma^\gamma$ and
\begin{equation}
\tc \sim \Sigma^{5 - 3 \gamma + (a/2) (3 - \gamma) + b (\gamma - 1)}.
\end{equation}
Table 1 gives a list of values for this scaling exponent for our nominal
value of $\gamma = 11/7$.  Notice that, when ice grains or metal grains
are evaporating, and in the bound-free/free-free opacity regime, cooling
time {\it decreases} as surface density {\it increases}.

Our cooling function is valid in the limit of large optical depth ($\tau
\gg 1$).  Since the disk becomes optically thin at some locations in the
course of a typical run, we must modify this result so that the cooling
rate does not diverge at small optical depth.  A modification that
produces the correct asymptotic behavior is
\begin{equation}\label{COOLNEW}
\Lambda = \frac{16}{3} \sigma T^4 \frac{\tau}{1 + \tau^2}.
\end{equation}
This interpolates smoothly between the optically-thick and
optically-thin regimes and is proportional to the (Rosseland mean)
optical depth in the optically-thin limit.  While it would be more
physically sensible to use a Planck mean opacity in the optically-thin
limit, usually the optically-thin regions contain little mass so their
cooling is not energetically significant.  An exception is in the
opacity gap, where even high density regions become optically thin.

Our simulations begin with $\Sigma$ and $U$ constant.  The velocity field is
perturbed from the equilibrium solution to initiate the gravitational
instability.  The initial velocities are $v_x = \delta v_x$, $v_y = -
{3\over{2}}\Omega x + \delta v_y$, where $\delta \bv$ is a Gaussian
random field of amplitude $\< \delta v^2 \>/c_s^2 = 0.1$. The
power spectrum of perturbations is white noise ($v_k^2 \sim k^0$) in a band
in wavenumber $k_{crit}/4 < |k| < 4 k_{crit}$ surrounding the minimum
$k_{crit} = 1/(\pi Q^2)$ (with $G = \Sigma_o = \Omega = 1$) in the
density-wave dispersion relation.  We have checked in particular cases
that for $10^{-3} < \< \delta v^2 \>/c_s^2 < 10$ the outcome is
qualitatively unchanged.  This is expected because disk
perturbations (unlike cosmological perturbations) grow exponentially and
the initial conditions are soon forgotten.

Excluding the initial velocity field, the initial conditions for a
spatially-uniform disk consist of three parameters: $\Sigma_o, U_o$, and $\Omega$.
We fix $Q = 1$, leaving two degrees of freedom.  In models with simple,
scale-free cooling functions such as that considered by \cite{gam01},
these degrees of freedom remain and can be scaled away by setting $G =
\Sigma_o = \Omega = 1$.  That is, there is a two-dimensional continuum
of disks (with varying values of $\Sigma_o$ and $\Omega$, but the same
value of $Q$) that are described by a single numerical
model.

The opacity contains definite physical scales in density and
temperature.  The realistic cooling function considered here therefore removes our freedom to
rescale the disk surface density and rotation frequency.  That is, there
is now a one-to-one correspondence between disks with fixed $\Sigma$ and
$\Omega$ and our numerical models.

The choice of $\Sigma_o$ and $\Omega$ as labels for the parameter space
is not unique.  Internally in the code we fix the initial
volume density (in $\gcm$) and the initial temperature (in
Kelvins).  These choices are difficult to interpret, however, since they
are tied to quantities that change over the course of the simulation;
$\Omega$ and the mean value of $\Sigma$ do not.

The cooling is integrated explicitly using a first-order scheme.  The timestep is
modified to satisfy the Courant condition and to be less than a fixed fraction of
the shortest cooling time on the grid. We have varied this fraction and shown that
the results are insensitive to it, provided that it is sufficiently small.

\section{Nonlinear Outcome}

\subsection{Standard Run}\label{STDRUN}

Consider the evolution of a single ``standard'' run, with $\Sigma_o =
1.4 \times 10^5 \gm \cm^{-2}$ and $\Omega = 1.1 \times 10^{-7}
\sec^{-1}$.  This corresponds to $T_o = 1200$ and $\tco = 9.0
\times 10^4 \Omega^{-1}$.  The model size is $L = 320
G\Sigma_o/\Omega^2$ and numerical resolution $1024^2$.  The model
initially lies at the lower edge of the opacity gap.

The evolution of the kinetic, gravitational and thermal energy per unit
area ($\<E_k\>$, $\<E_g\>$ and $\<E_{th}\>$ respectively) normalized to
$G^2\Sigma_o^3/\Omega^2$,\footnote{The natural unit that can be formed
from $G$, $\Sigma$ and $\Omega$.} are shown in Figure 1.  After the initial phase
of gravitational instability the model settles into a statistically-steady,
gravito-turbulent state.  It does not fragment.  Cooling is
balanced by shock heating.  Energy for driving the shocks is extracted
from the shear flow, and the mean shear flow is enforced by the boundary
conditions.

The turbulent state transports angular momentum outward via hydrodynamic
and gravitational shear stresses.  The dimensionless gravitational shear stress
is
\begin{equation}
\alpha_{grav} = \frac{1}{\<\frac{3}{2} \Sigma c_s^2\>}
		\int_{-\infty}^{\infty} dz {g_x g_y\over{4 \pi G}}
\end{equation}
where ${\bf g}$ is the gravitational acceleration, and
the dimensionless hydrodynamic shear stress is
\begin{equation}
\alpha_{hyd} = \frac{\Sigma v_x \delta v_y}{\<\frac{3}{2} \Sigma c_s^2\>}
\end{equation}
where $\<\>$ denote a spatial average. Figure 2 shows the evolution of $\< \alpha_{grav} \>$
and $\< \alpha_{hyd} \>$ in the standard run.  Averaged over the last $230 \Omega^{-1}$
of the run, $\<\< \alpha_{hyd} \>\> = 0.0079$, $\<\< \alpha_{grav} \>\> = 0.017$, and so the
total dimensionless shear stress is $\<\< \alpha \>\> = 0.025$, where $\<\<\>\>$ denote
a space and time average.

The mean stability parameter $\<Q\> \equiv \langle c_s
\rangle\Omega/\pi G\langle \Sigma\rangle$ averages $1.86$ over the
last $230 \Omega^{-1}$ of the run.  Because the temperature and
surface density vary strongly, other methods of averaging $Q$ will
give different results.

Figure 3 shows a snapshot of the surface density at $t = 50
\Omega^{-1}$.  The structure is similar to that observed in
\cite{gam01}, with trailing density structures.  The density
structures are stretched into a trailing configuration by the
prevailing shear flow.  Their scale is determined by the disk
temperature and surface density rather than the size of the box (see
\citealt{gam01}).

\subsection{Varying $\Sigma_o$ and $\Omega$}

We now turn to exploring the two-dimensional parameter space of
models.  First consider a series of models with the same initial
central temperature, but with varying $\tco$.  As $\tco$ is lowered
the time-averaged gravitational potential energy per unit area $\<\<E_g\>\>$ increases
monotonically in magnitude.  The gravito-turbulent state becomes more
extreme, with larger $\<\< \alpha \>\>$, larger perturbed velocities, and larger
density contrasts.  Eventually a threshold is crossed and the disk
fragments.

Fragmentation is illustrated in Figure 4, which shows a snapshot from a
run with $\Sigma_o = 6.6 \times 10^3 \gm \cm^{-2}$, $\Omega = 5.4 \times
10^{-9} \sec^{-1}$.  This corresponds to $T_o = 1200$,  $\tco =
0.025\Omega^{-1}$.  The run has numerical resolution $256^2$ and $L = 80
G\Sigma_o/\Omega^2$.  The largest bound object in the center of the
figure was formed from the collision and coalescence of several smaller
bound objects.  A snapshot of the optical depth at the same point in the
simulation is given in Figure 5.  For each snapshot, red indicates high
values of the mapped variable and blue indicates low values.  Much of
the disk is optically thick, but most of the low density regions are
optically thin in the Rosseland mean sense.

Lowering $\tco$ sufficiently always leads to fragmentation.  We have
surveyed the parameter space of $\Omega$ and $\Sigma_o$ to
determine where the disk begins to fragment.  Each model was run to $100
\Omega^{-1}$.\footnote{In four cases we had to run the simulation longer
to get converged results.}  Figures 6 and 7 summarize the results.  Two heavy solid
lines are shown on each diagram.  The upper line shows the most rapidly
cooling simulations that show no signs of gravitational fragmentation
({\it nonfragmentation point}).  Quantitatively, we define this as the
point at which the time-averaged gravitational potential energy per unit
area is equal to $-3 G^2 \Sigma_o^3/\Omega^2$.\footnote{$-3$ is the potential
energy per unit area of a wave at the critical wavelength in a $Q = 1$
disk with $\delta \Sigma/\Sigma = \sqrt{3}/\pi$. No bound objects are
observed throughout the duration of these runs.} The lower line shows
the most slowly cooling simulations to show definite fragmentation ({\it
fragmentation point}).  Quantitatively, we define this as the point at
which the gravitational potential energy per unit area is equal to $-300
G^2 \Sigma_o^3/\Omega^2$ {\it at some point during the run}.\footnote{These
runs exhibit bound objects that persist for the duration of the run.}  Figure 6 shows the data
in the $\rho_o, T_o$ plane, while Figure 7 shows the results in the $\Sigma_o,
\Omega$ plane. The light contours are lines of constant $\tco$.

The transition from persistent, gravito-turbulent outcomes to
fragmentation is gradual and statistical in nature.  Figure 8 shows the
gravitational potential energy per unit area in the transition region for a series
of runs with $T_o = 1200\K$.  The abscissa is labeled with the initial
cooling time $\tco\Omega$.  There is a gradual, approximately
logarithmic increase in the magnitude of $\<\<E_g\>\>$ as $\tco$ decreases.
Runs in this region exhibit the transient formation of small bound
objects, which might collapse if additional physics (e.g.  the effects of
MHD turbulence) were included in the model.  Eventually $-\<\<E_g\>\>$ begins to
increase dramatically, and we define the {\it transition point} as the
beginning of this steep increase in gravitational binding energy.

Figure 9 shows the run of $\tco\Omega$ for the fragmentation point,
transition point, and nonfragmentation point as a function of $T_o$.
It is surprising that a disk can begin to exhibit signs of
gravitational collapse for $\tco\Omega$ as large as $10^6$, and evade
collapse for $\tco\Omega$ as small as $0.02$.  A naive application of
the results of \cite{gam01} would suggest that fragmentation should
occur for $\tco\Omega \lesssim 3$.  Evidently this estimate can be off by
orders of magnitude, with the largest error for $T_o \approx 10^3\K$,
just below the opacity gap.

The physical argument for fragmentation at short cooling times is as
follows (e.g. \citealt{shlos90}).  Thermal energy is supplied to the
disk via shocks.  Strong shocks occur when dense clumps collide with one
another; this occurs on a dynamical timescale $\sim \Omega^{-1}$.  If
the disk cools itself more rapidly then shock heating cannot match
cooling and fragmentation results.  This argument is apparently
contradicted by Figure 9.  The resolution lies in finding an appropriate
definition of cooling time.  The disk loses thermal energy on the
effective cooling timescale
\begin{equation}
\tce^{-1} \equiv {\<\< \Lambda \>\> \over{\<\< U \>\>}}.
\end{equation}
Figure 10 shows the run of
$\tce$ at the fragmentation, transition, and non-fragmentation points.
Evidently $\tce$ at transition lies between $\Omega^{-1}$ and $10
\Omega^{-1}$.  Figure 11 shows the run of $\tco$ and $\tce$ on the
transition line.  Just below the opacity gap they differ by as much as
four orders of magnitude.

Why do $\tco$ and $\tce$ differ by such a large factor?  The answer is
related to the existence of sharp variations in opacity with
temperature.  Consider a disk near the lower edge of the opacity gap.
Once gravitational instability sets in, fluctuations in temperature move
parts of the disk into the opacity gap.  There, the opacity is reduced
by orders of magnitude.  Since the cooling rate for an optically thick
disk is proportional to $\kappa^{-1}$, the cooling time drops by a
similar factor.   Relatively small variations in temperature can thus
produce large variations in cooling rate.

As in \cite{gam01}, the result $\tce\Omega \gtrsim 1$ also implies a
constraint on $\<\< \alpha \>\>$.  Energy conservation implies that
\begin{equation}\label{AVGSOL}
\qq \Omega \<\< W_{xy} \>\> = \<\< \Lambda \>\>,
\end{equation}
where $W_{xy}$ is the total shear stress (hydrodynamic plus gravitational). Equivalently, stress by rate-of-strain
is equal to the dissipation rate. Using the definition of $\tce$, this implies
\begin{equation}\label{ANALPHA}
\<\< \alpha \>\> = \left(\gamma (\gamma - 1) {9\over{4}} \Omega \tce
	\right)^{-1}.
\end{equation}
Hence $\tce\Omega \gtrsim 1$ implies $\<\< \alpha \>\> \lesssim 1$.  Figure 12
shows $\<\< \alpha \>\>$ vs $\tce$ for a large number of runs plotted against
equation (\ref{ANALPHA}).  For small values of $\tce$ the numerical
values lie below the line.  These models are not in equilibrium (i.e., not in a
statistically-steady gravito-turbulent state), so the
time average used in equation (\ref{AVGSOL}) is not well defined.  For
larger values of $\tce$ numerical results typically (there is noise in
the measurement of both $\<\< \alpha \>\>$ and $\tce$ because the time
average is taken over a finite time interval) lie slightly above the
analytic result.

The bias toward points lying slightly above the line reflects the fact
that $\<\< \alpha \>\>$ measures the rate of energy extraction from the shear
while $\tce$ measures the rate at which that energy is transformed into
thermal energy.  If energy is lost, perhaps to numerical averaging at
the grid scale, then more energy must be extracted from the shear flow
to make up the difference.  Overall, however, the agreement with the
analytic result is good and demonstrates good energy conservation
in the code.

The relationship between $\tce$ and $\<\< \alpha \>\>$ is interesting but not
particularly useful because $\tce$ is no more readily calculated than
$\<\< \alpha \>\>$; it depends on a complicated moment of the surface density and
temperature.  Only for constant cooling time have we been able to evaluate this moment
analytically.

\subsection{Isothermal Disks}

We have assumed that external illumination of the disk is negligible.
This approximation is valid when the effective temperature $T_{irr}$ of
the external irradiation is small compared to the central temperature of
the disk.  In the opposite limit, illumination controls the energetics
of the disk and it is isothermal (if it is illuminated directly so that
shadowing effects, such as those considered by \cite{cond03} are
negligible).

It is therefore worth studying the outcome of gravitational instability
in an isothermal disk.  The isothermal disk model has a single
parameter: the initial value of $Q$.  We ran models with varying values
of $Q$ and with $\<\delta v^2\>/c_s^2 = 0.1$.  We find that models with
$Q \lesssim 1.4$ fragment.

It is likely that the mass of the fragments, etc., depends on how an
isothermal disk becomes unstable.  Rapid fluctuation of the external
radiation field is likely to produce a different outcome than dimming on a
timescale long compared to the dynamical time.

\section{Discussion}

Using numerical experiments, we have identified
those disks that are likely to fragment absent external
heating.  Disks with effective cooling times $\tce \lesssim
\Omega^{-1}$ are susceptible to fragmentation.  This is what one might
expect based on the simple argument of \cite{shlos90}: if the disk cools
more quickly than the self-gravitating condensations can collide with
one another, then those collisions (which occur on a timescale $\sim
\Omega^{-1}$) cannot reheat the disk and fragmentation is inevitable.  But our
results are at the same time surprising.

The effective cooling time depends on the nonlinear outcome of
gravitational instability.  It depends on the cooling function, which in
turn depends sensitively on $\Sigma$ and $U$.  Since $\Sigma$ and $U$
vary strongly over the disk once gravitational instability has set in,
it is difficult to estimate $\tce$ directly.  One might be tempted to
estimate $\tce(\Sigma,\Omega) \simeq \tco(\Sigma_o,\Omega,Q = 1)$, but
our experiments show that this estimate can be off by as much as four
orders of magnitude.  The effect is particularly pronounced near sharp
features in the opacity.  For example, consider a model initially
located just below the opacity gap with $\tco\Omega \gg 1$.
Gravitational instability creates dense regions with higher
temperatures, where dust is destroyed.  The result is rather like having
to shed one's blanket on a cold winter morning: the disk loses its
thermal energy suddenly.  Pressure support is lost and gravitational
collapse ensues.

The difference between $\tce$ and $\tco (Q = 1)$ implies that a much
larger region of the disk is susceptible to fragmentation than naive
estimates based on the approximation $\tce \approx \tco$ might suggest.
For example, consider an equilibrium disk model with $Q \gg 1$ at small
$r$.  As $r$ increases, $Q$ declines.  Eventually $Q\sim 1$ and
gravitational instability sets in.  There is then a range of radii where
$Q \sim 1$, $\tce\Omega \gtrsim 1$ and recurrent gravitational
instability can transport angular momentum and prevent collapse.
Generally speaking, however, the cooling time decreases with increasing
radius.  Eventually  $\tce\Omega \sim 1$ and fragmentation cannot be
avoided.  By lowering our estimate of $\tce$, we narrow the range of
radii over which recurrent gravitational instability can occur.

The general sense of our result is that it is extremely difficult to
prevent a marginally-stable, $Q \sim 1$, optically-thick disk from
fragmenting and forming planets (in circumstellar disks) or stars (in
circumstellar and circumnuclear disks).  This is particularly true for
disks with $T \sim 10^3\K$, whose opacity is dominated by dust grains,
i.e. disks whose temperature lies within a factor of several of the
opacity gap.

Our numerical model uses a number of approximations.  First, our
treatment is razor-thin, i.e. all the matter is in a thin slice at
$z = 0$.  The effect of finite thickness on linear stability has been
understood since \cite{glb}: it is stabilizing because gravitational
attraction of neighboring columns of disk is diluted by finite
thickness.  The size of the effect may be judged by the fact that $Q =
0.676$ is required for marginal stability of a finite-thickness,
isothermal disk.  

The behavior of a finite-thickness disk in the nonlinear regime is more
difficult to predict.  Shocks will evidently deposit some of their
energy away from the midplane, where it can be radiated away more
quickly (because the energy is deposited at smaller optical depth - see \citealt{pic00}).
Radiative diffusion parallel to the disk plane (not included here) may
enhance cooling of dense, hot regions.  Both these effects are
destabilizing.  Ultimately, however, a numerical study is required.
This is numerically expensive:  one
must resolve the disk vertically, on the scale height $H$, and
horizontally, at the critical wavelength $2 \pi Q H$.

Second, we have ignored magnetic fields.  While there may be
astrophysical situations where cool disks have such low ionization that
they are unmagnetized, most disks are likely to contain dynamically
important magnetic fields that give rise to a dimensionless shear stress
$\<\< \alpha \>\> \gtrsim 0.01$ (e.g. \citealt{hgb1}).  These fields are likely to
remove spin angular momentum from partially collapsed objects,
destabilizing them.  Numerical experiments including both gravitational
fields and magnetohydrodynamics are necessarily three dimensional (the
instability of \cite{bh91} requires $\partial_z \ne 0$), and are thus
numerically expensive.

Third, we have fixed $\gamma$ and $\mu$ for the duration of each
simulation.  This eliminates the soft spots in the equation of state
associated with ionization of atomic hydrogen and dissociation of
molecular hydrogen.  In these locations the three dimensional $\gamma$
dips below $4/3$, which is destabilizing.

Fourth, we have treated the physics of grain destruction and formation
very simply.  In using the \cite{bel94} opacities we implicitly assume
that grains reform in cooling gas on much less than a dynamical time.
It is likely that grain re-formation will take some time (e.g.
\citealt{hes91}) and this will further reduce the disk opacity and
enhance fragmentation.

Fifth, we have neglected the effects of illumination.  In the limit of
strong external illumination, i.e. when the effective temperature of the
irradiation $T_{irr}$ is large compared to the disk central temperature
$T_c$, the disk is isothermal (here $T_c$ is the temperature of a dense
condensation).  We have carried out isothermal experiments and shown
that, for initial velocity perturbations with $\< \delta v^2 \>/c_s^2 = 0.1$,
disks with $Q \lesssim 1.4$ fragment.  Weaker illumination produces a more
complicated situation that we have not explored here.
Illumination-dominated disks that become unstable presumably do so
because the external illumination declines, and the rate at which the
external illumination changes may govern the nonlinear outcome.

We conclude that disks with $\tce\Omega \lesssim 1$ do not exist.
Cooling in this case is so effective that fragmentation into condensed
objects-- stars, planets, or smaller accretion disks-- is inevitable.

As an example application of this result, consider the model for the
nucleus of NGC 1068 recently proposed by \cite{lb03}.  Their model is an
extended marginally-stable self-gravitating disk of the type
investigated here and originally proposed by \cite{glb} for galactic
disks and \cite{pac78} for accretion disks, although their disk is
sufficiently massive that it modifies the rotation curve as well.  Based
on their Figure 3, at a typical radius of $0.5\pc$, $\Sigma_o \simeq
10^4$ and $\Omega \simeq 10^{-9}$.  According to our Figure 7 this disk is
about 2 orders of magnitude too dense to avoid fragmentation.  While it
may be possible to avoid this conclusion by invoking strong external
heating, the energy requirements are severe, as outlined in
\cite{goo03}.  The disk proposed by \cite{lb03} would therefore fragment
into stars on a short timescale.

This work was supported by NSF grant AST 00-03091, PHY 02-05155, and
NASA grant NAG 5-9180.

\newpage

\begin{deluxetable}{lccc}
\tablenum{1}
\tablecolumns{4}
\tablewidth{0pc}
\tabcolsep 0.5truecm
\tablecaption{Scaling Exponent for Cooling Time as a Function of Surface Density}
\tablehead{Opacity Regime & a & b & Exponent}
\startdata
Ice grains & 0 & 2 & 10/7 \\
Evaporation of ice grains & 0 & -7  & -26/7 \\
Metal grains & 0 & 1/2  & 4/7 \\
Evaporation of metal grains & 1 & -24  & -89/7 \\
Molecules & 2/3 & 3  & 52/21 \\
H$^-$ scattering & 1/3 & 10  & 131/21 \\
Bound-free and free-free & 1 & -5/2  & -3/7 \\
Electron scattering & 0 & 0  & 2/7 \\
\enddata
\end{deluxetable}

\begin{figure}
\plotone{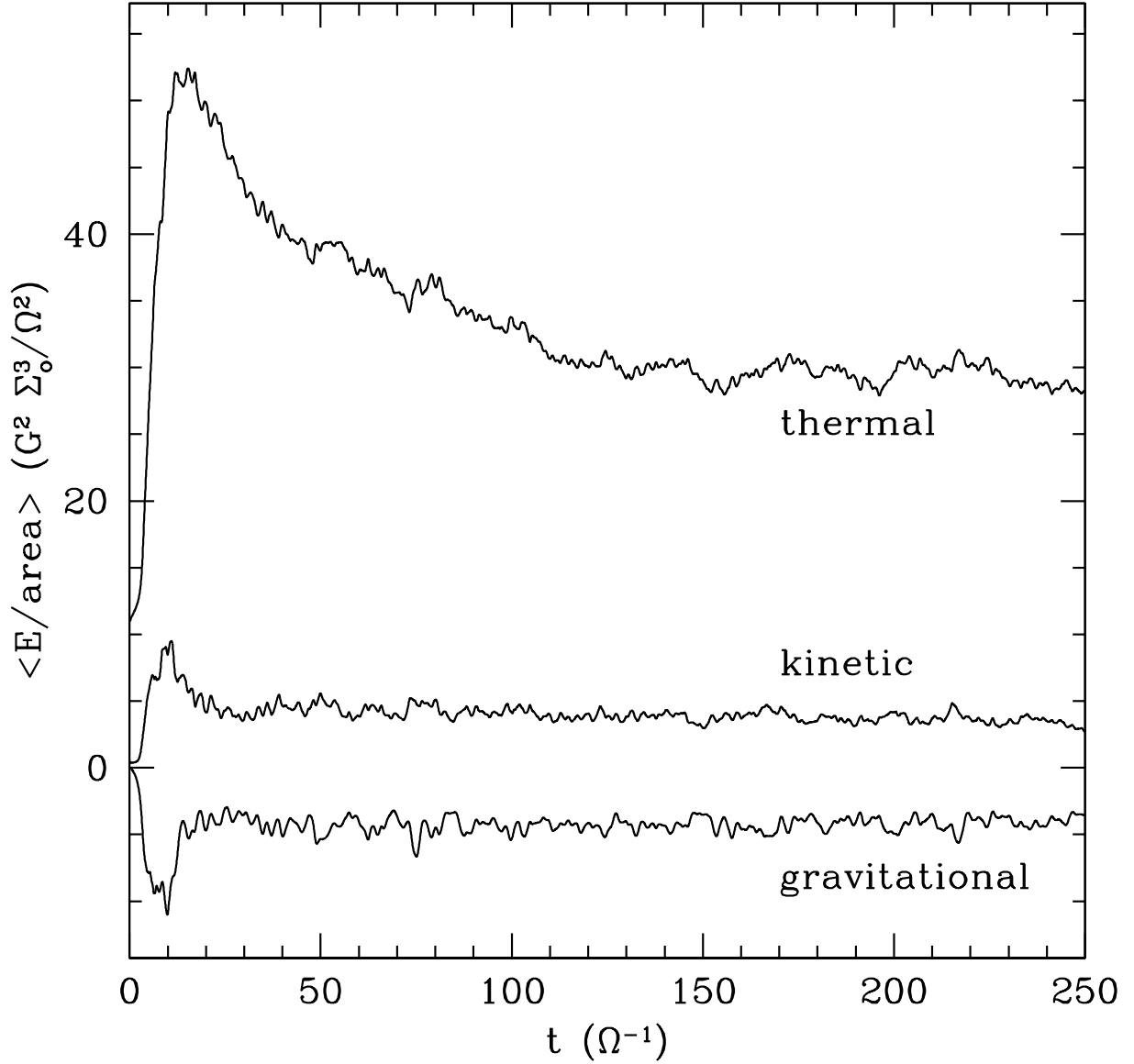}
\caption{
Evolution of the kinetic, gravitational, and thermal energy per unit
area, normalized to $G^2 \Sigma_o^3/\Omega^2$, in the standard run, which
has $L = 320 G\Sigma_o/\Omega^2$, resolution $1024^2$, and $\tco = 9.0
\times 10^4 \Omega^{-1}$.
}
\end{figure}

\begin{figure}
\plotone{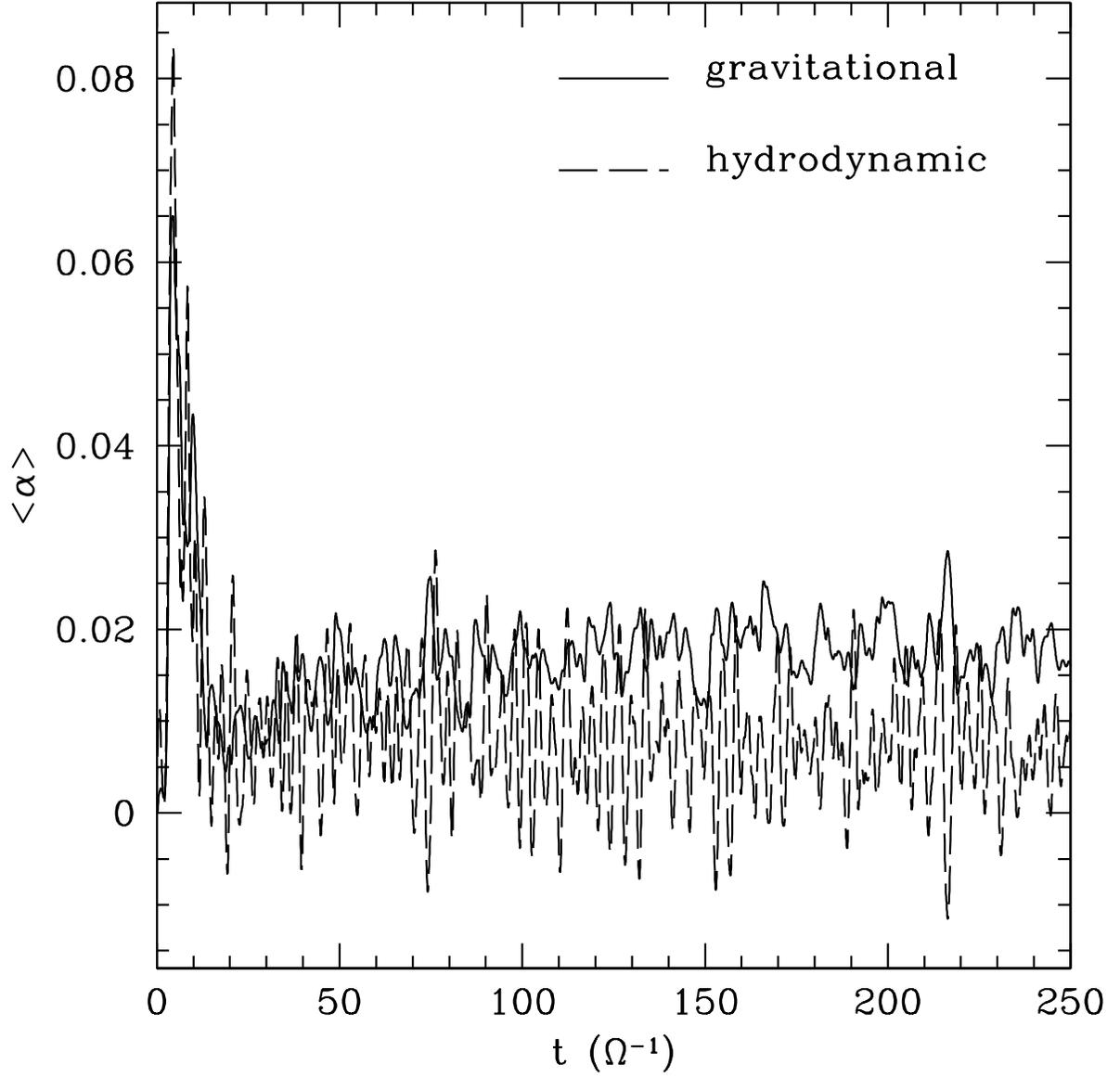}
\caption{
Evolution of the gravitational (solid line) and hydrodynamic (dashed
line) pieces of $\< \alpha \>$ in the standard run.
}
\end{figure}

\begin{figure}
\plotone{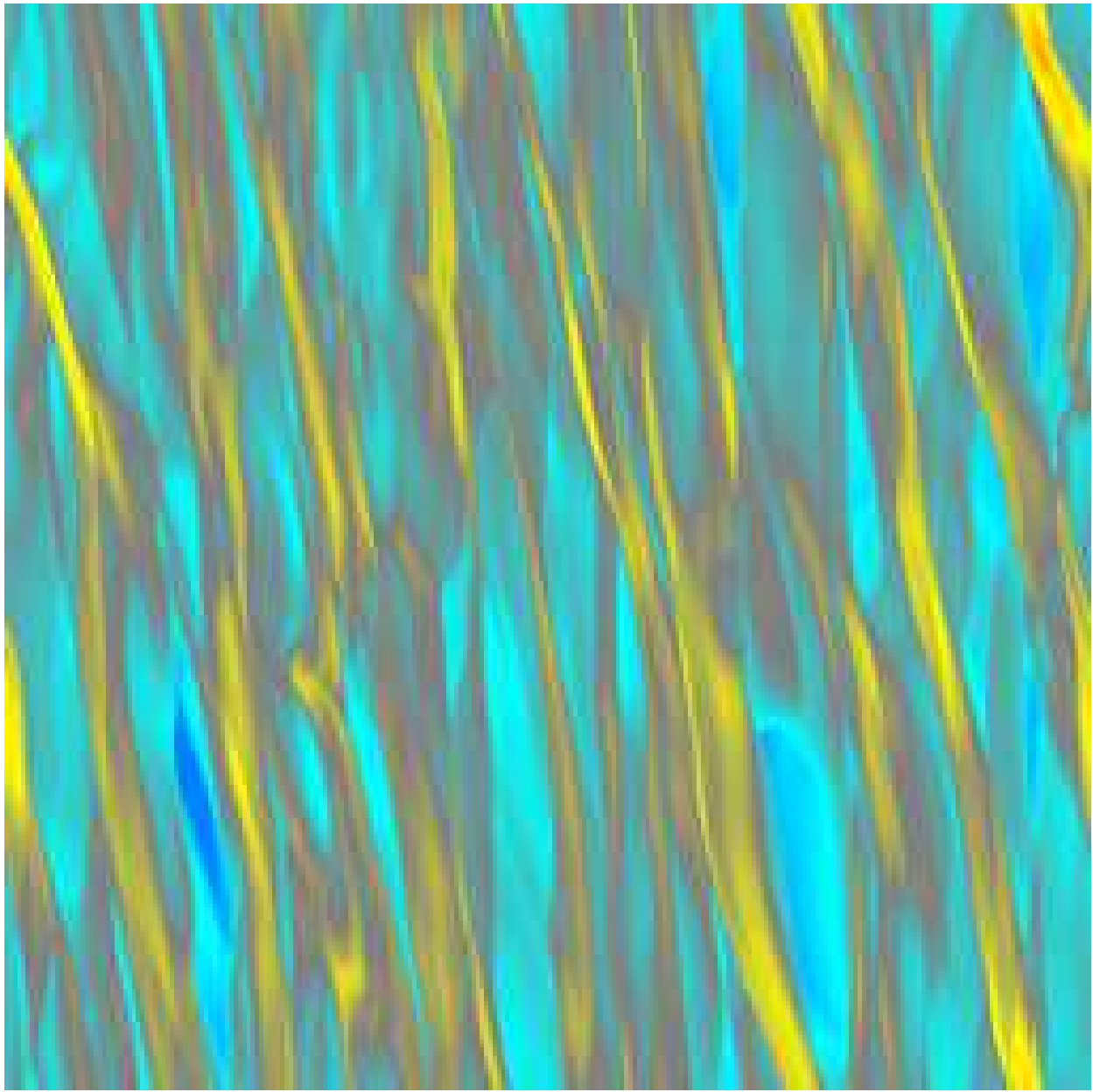}
\caption{
Map of surface density at $t = 50 \Omega^{-1}$ in the standard run.
Blue is low density ($0.2 \Sigma_o$) and yellow is high density ($3
\Sigma_o$).
}
\end{figure}

\begin{figure}
\plotone{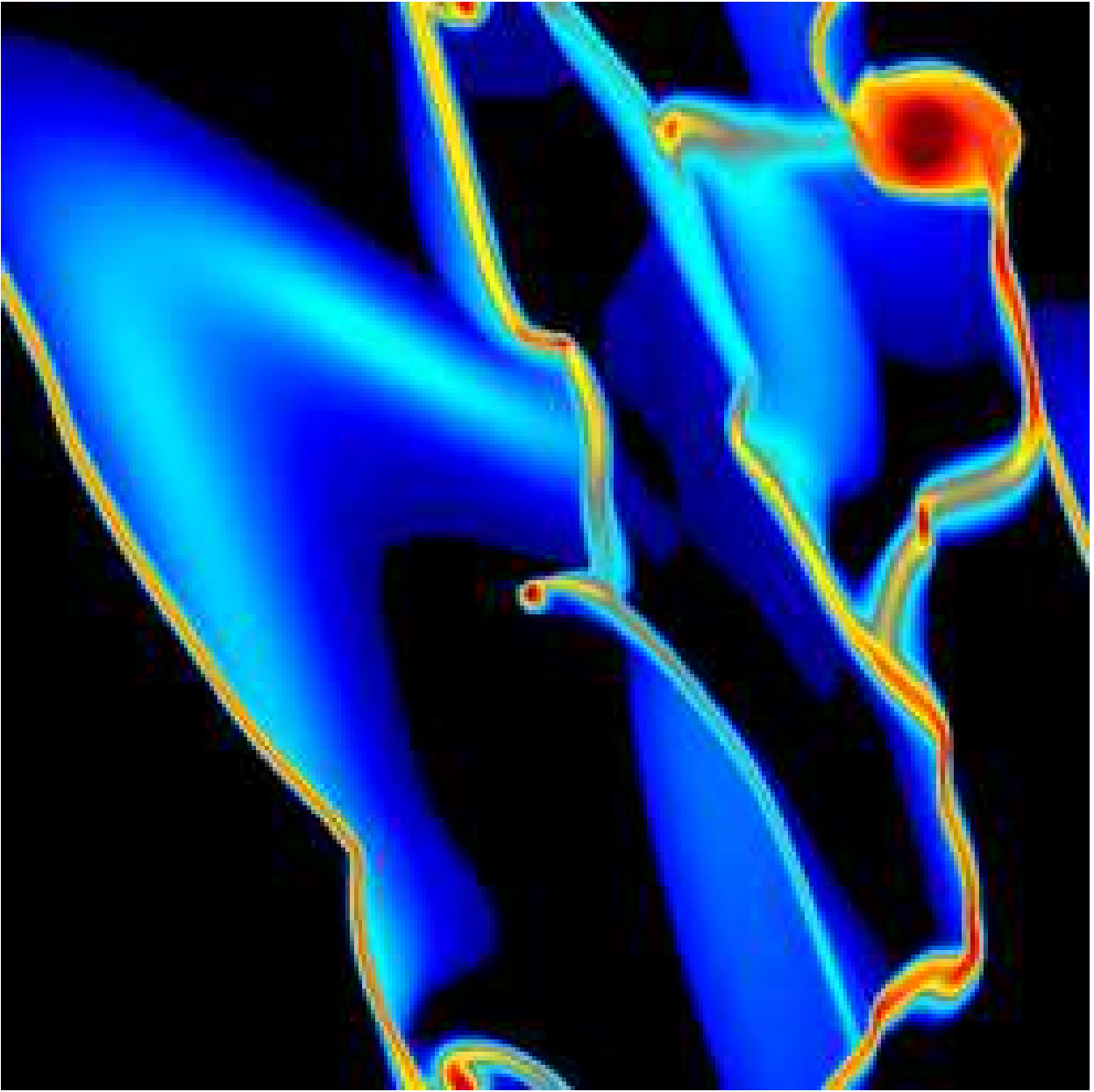}
\caption{
Map of surface density in a run with $\tco = 0.025\Omega^{-1}$.  Black is
low density ($10^{-2} \Sigma_o$) and red is high density ($10^2
\Sigma_o$).
}
\end{figure}

\begin{figure}
\plotone{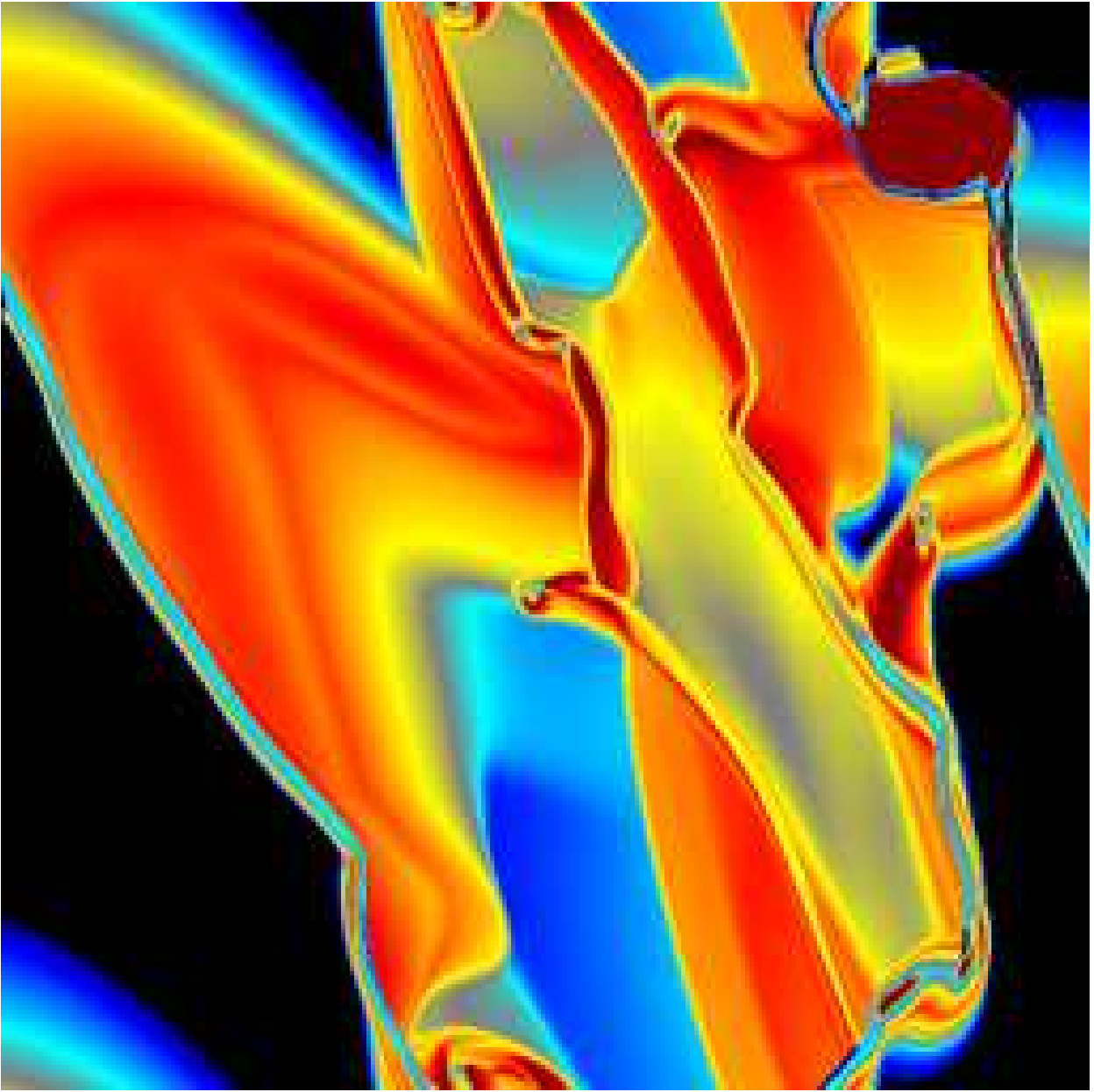}
\caption{
Map of optical depth in a run with $\tco = 0.025 \Omega^{-1}$. Black is low
optical depth ($10^{-2}$) and red is high optical depth ($10^4$).
}
\end{figure}

\begin{figure}
\plotone{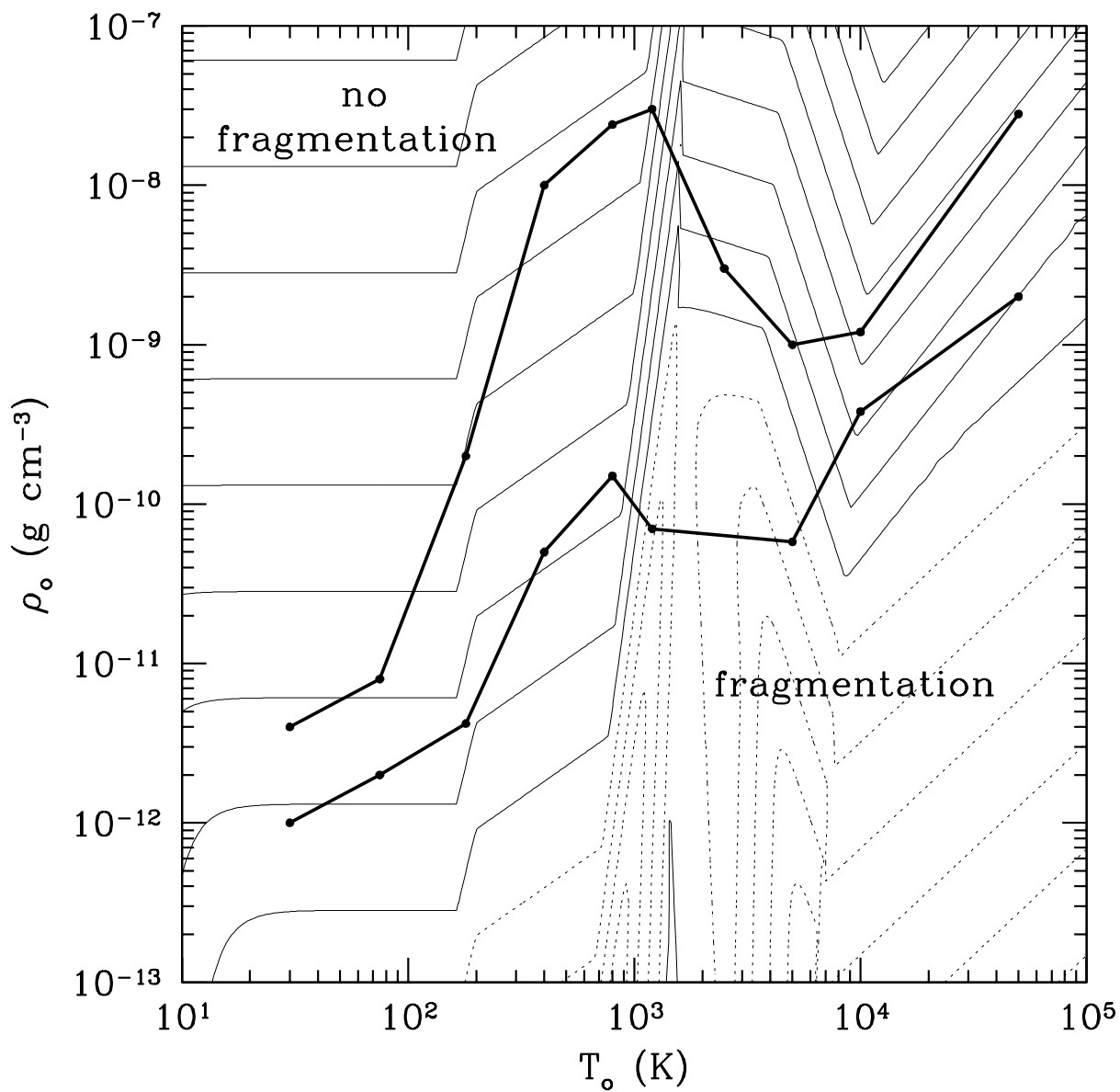}
\caption{
Location of the critical curves as a function of initial volume density and
temperature (in cgs units). Each contour line is an order of magnitude
change in $\tco$, solid/dotted lines indicating positive/negative
integer values of log($\tco$).
} \end{figure}

\begin{figure}
\plotone{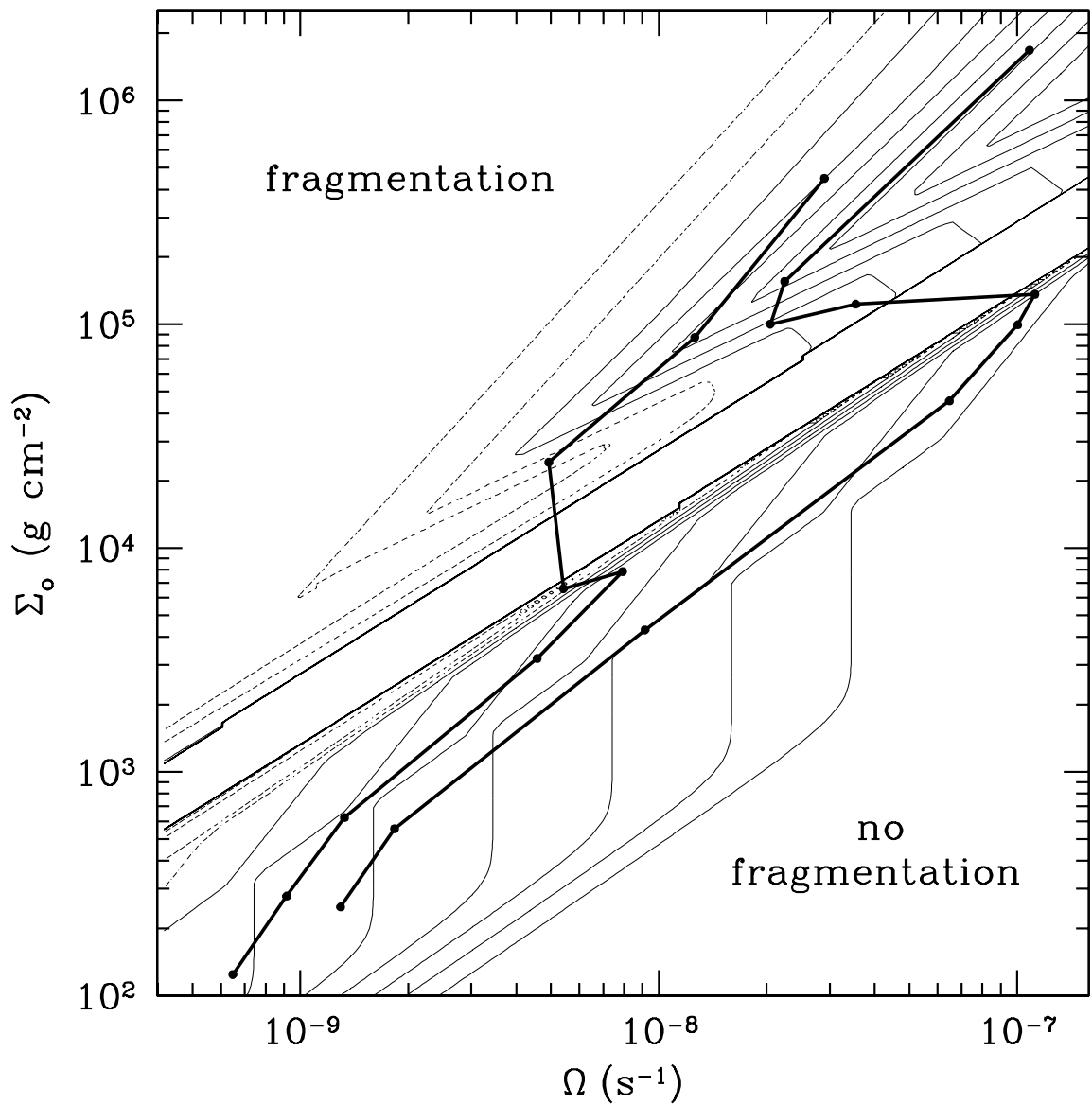}
\caption{
Location of the critical curves as a function of initial surface density
and rotation frequency (in cgs units).  Each contour line is an order of
magnitude change in $\tco$, solid/dotted lines indicating
positive/negative integer values of log($\tco$).  The gap in the center
of the plot is due to the discontinuous jump in the value of $\mu$.
}
\end{figure}

\begin{figure}
\plotone{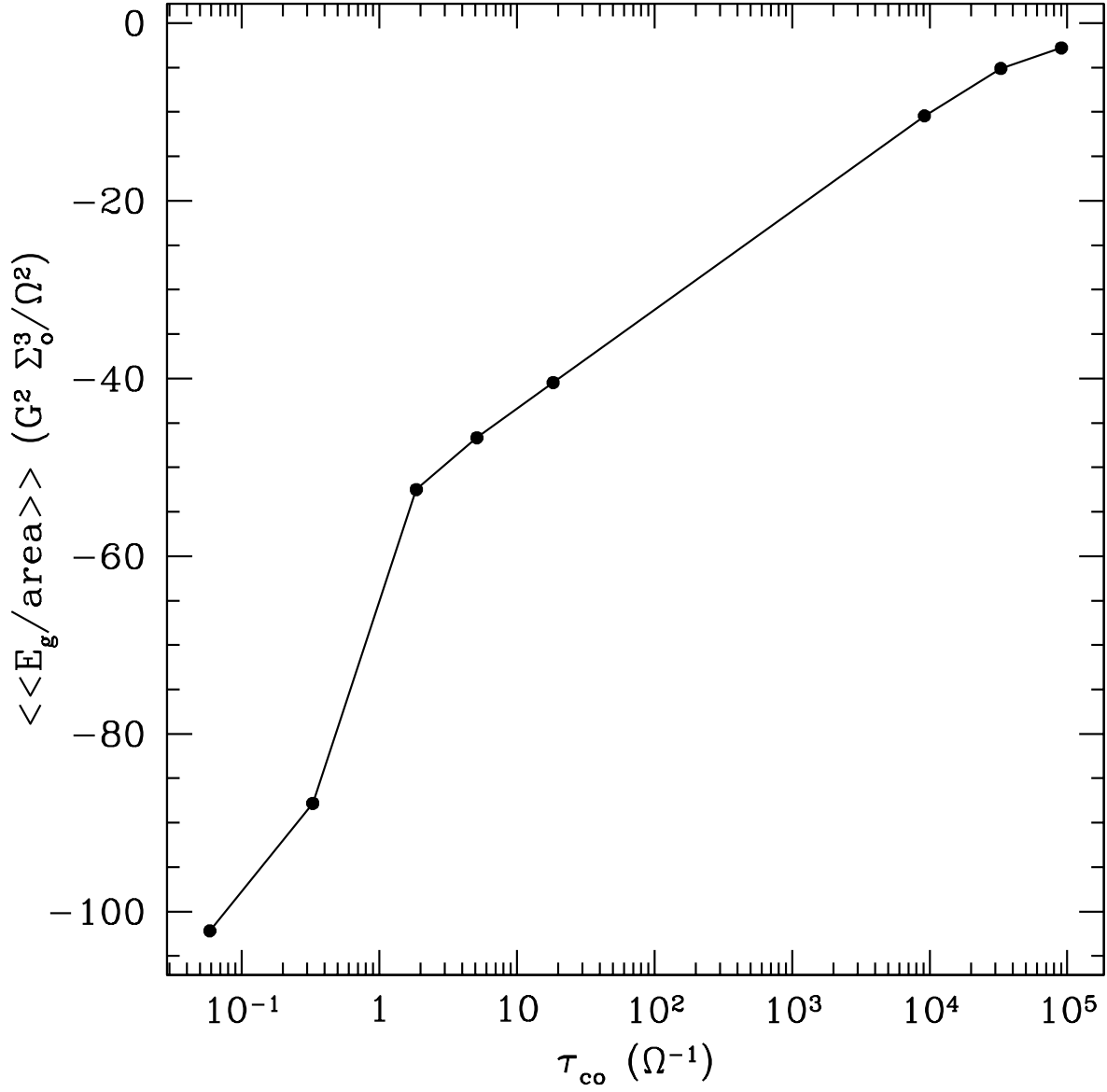}
\caption{
Mean gravitational potential energy as a function of initial cooling time
for a series of models with varying initial cooling time and $T_o =
1200$. 
}
\end{figure}

\begin{figure}
\plotone{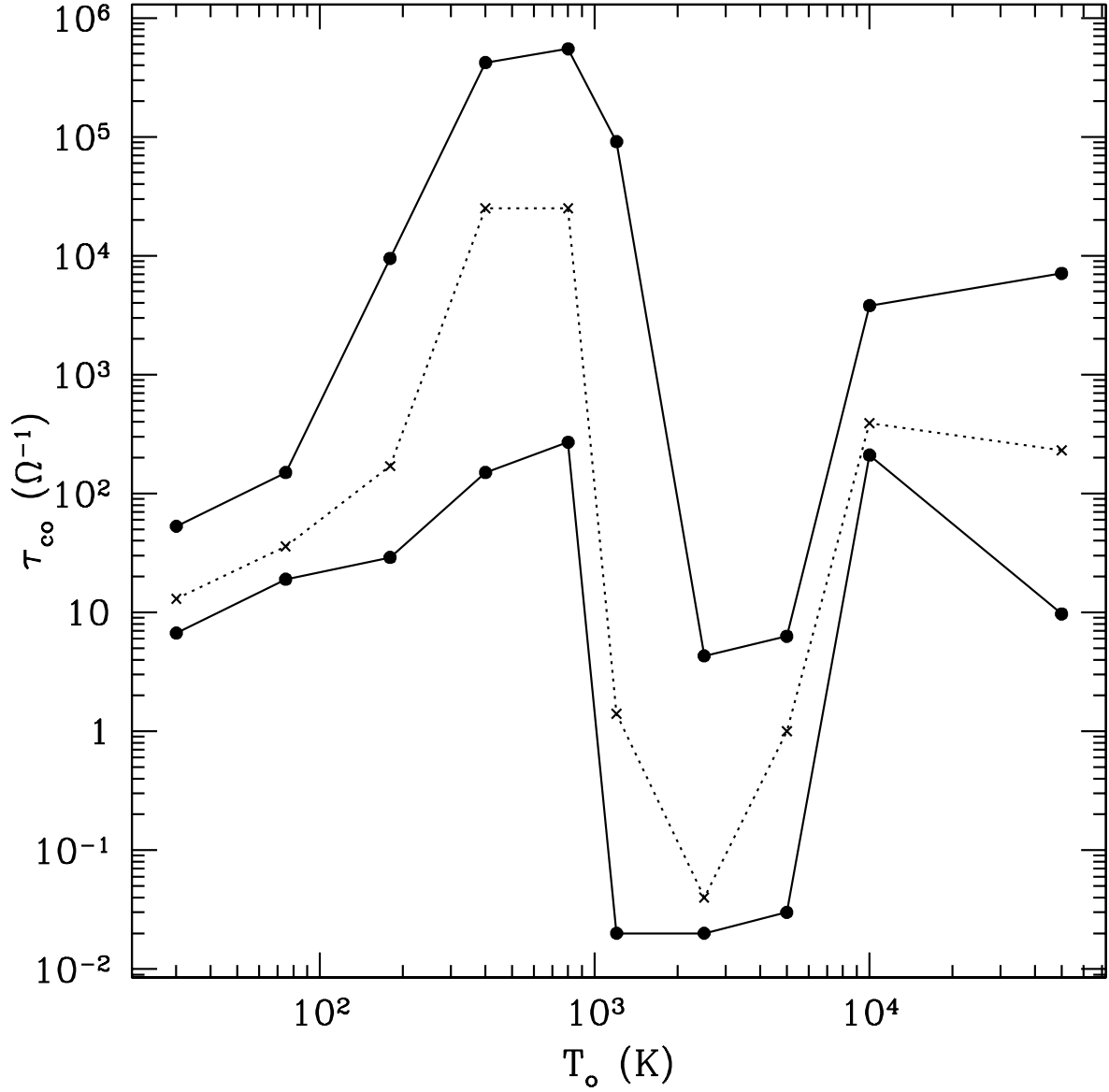}
\caption{
Initial cooling times at the points of non-fragmentation, fragmentation
and transition. 
}
\end{figure}

\begin{figure}
\plotone{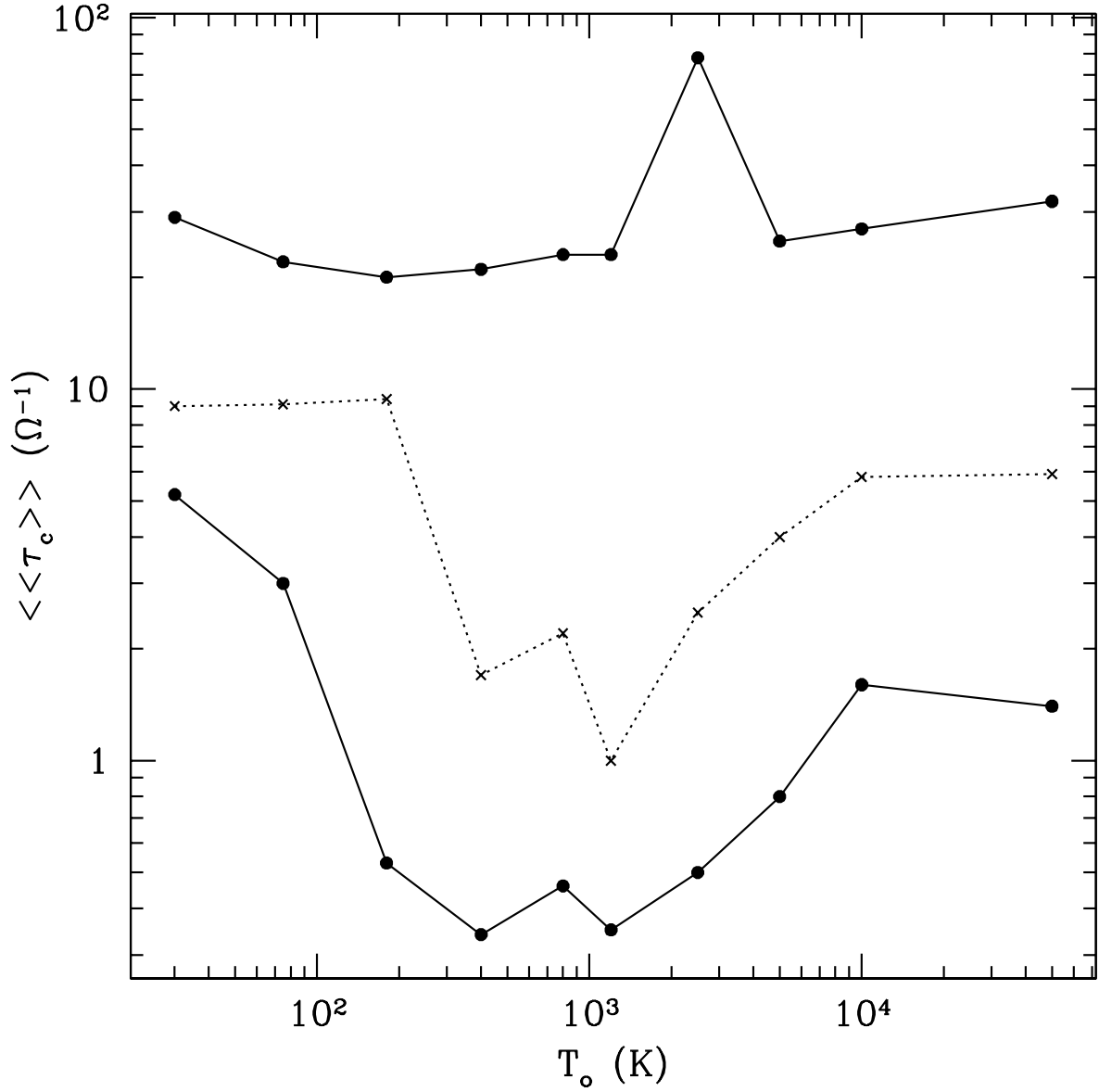}
\caption{
Effective cooling times at the points of non-fragmentation,
fragmentation and transition. 
}
\end{figure}

\begin{figure}
\plotone{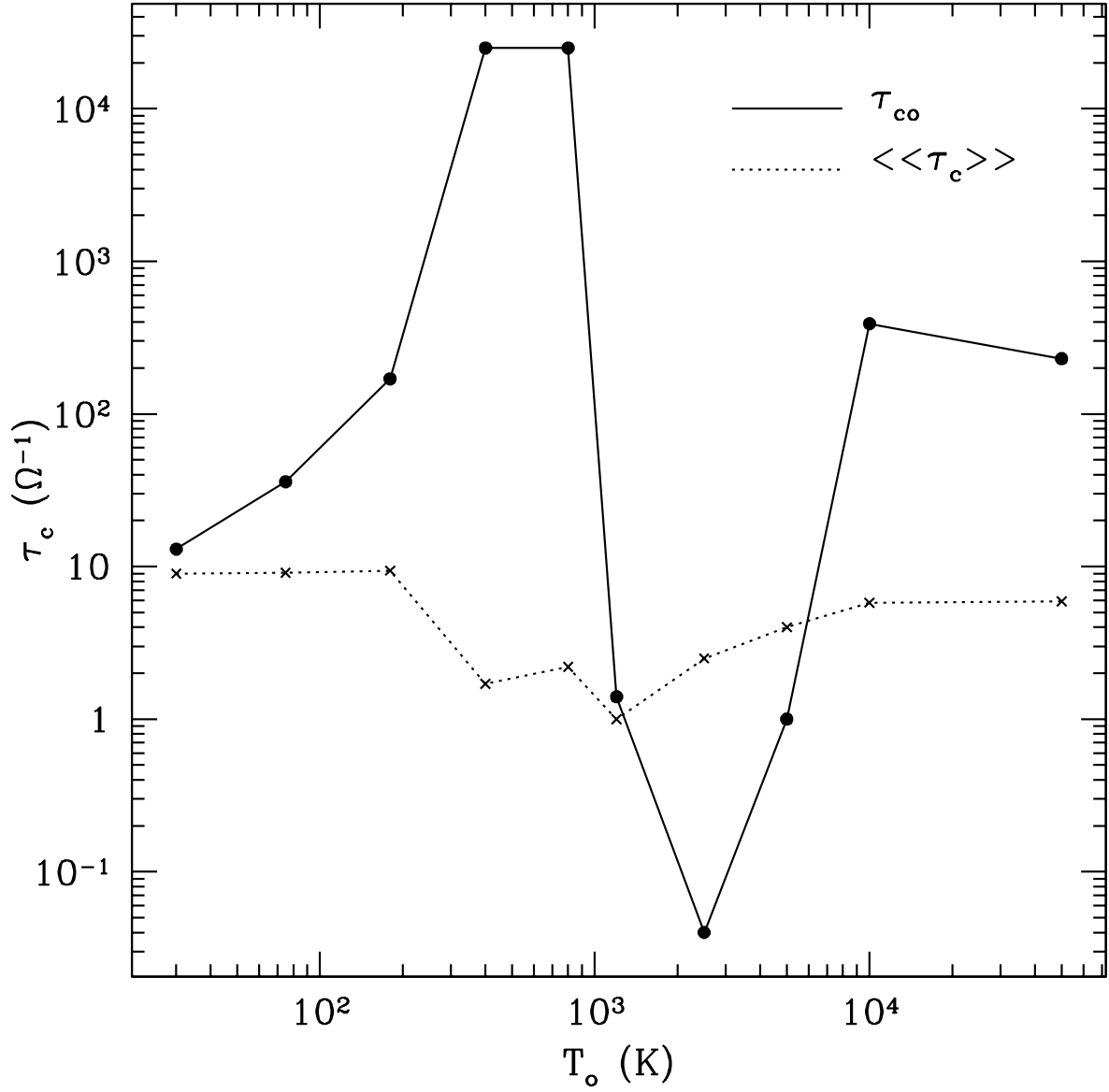}
\caption{
Initial and effective cooling times at the transition between
non-fragmentation and fragmentation. 
}
\end{figure}

\begin{figure}
\plotone{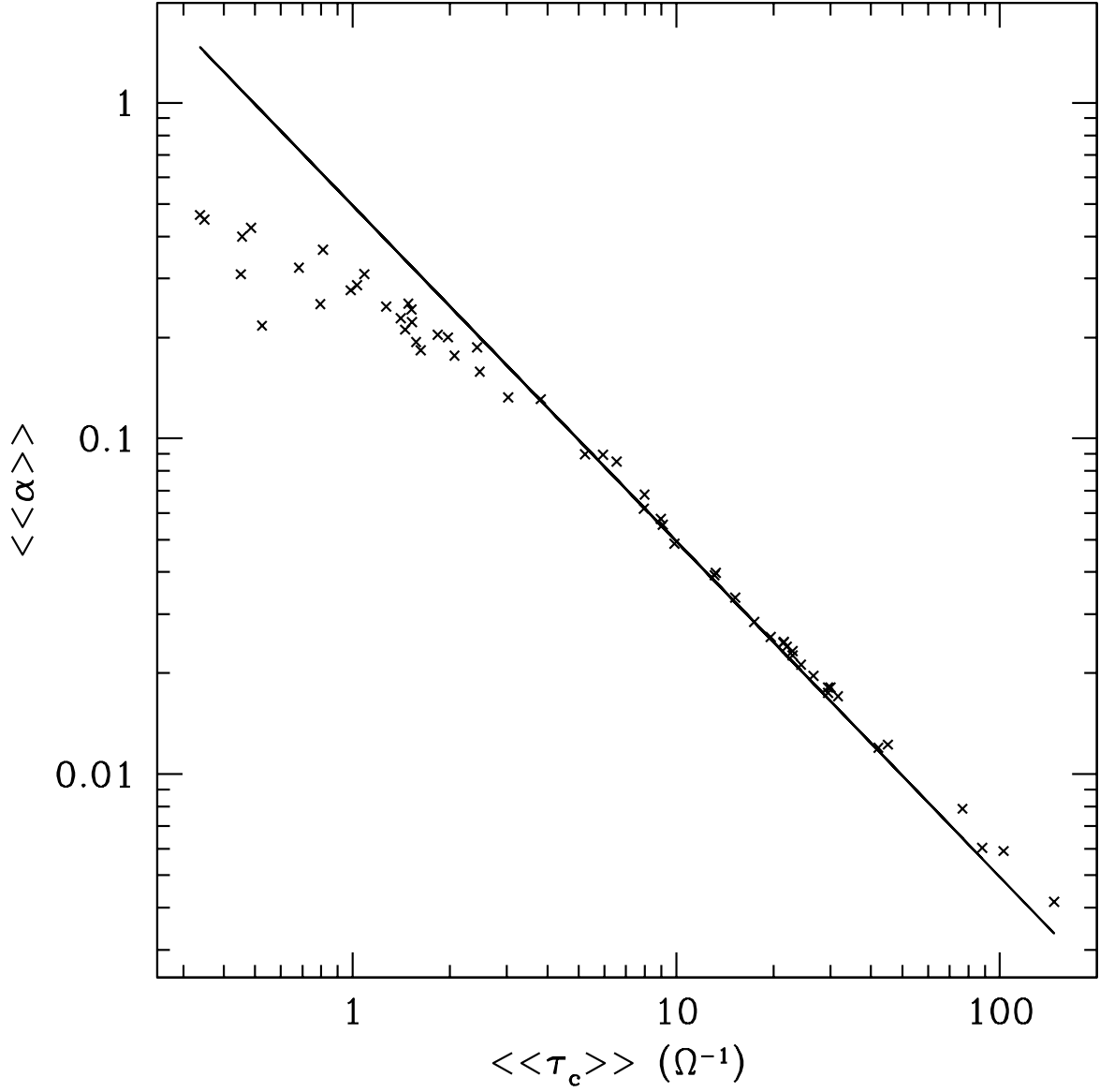}
\caption{
Time-averaged shear stress vs. effective cooling time for a series of
runs. The solid line shows the analytic result, based on energy
conservation, from equation (\ref{ANALPHA}).
} \end{figure}

\end{document}